# N-polar AlN buffer growth by MOVPE for transistor applications

Jori Lemettinen[1]*, Hironori Okumura[2,3], Tomás Palacios[3], and Sami Suihkonen[1]

*[1]Department of Electronics and Nanoengineering, Aalto University, P.O BOX 13500, FIN-00076 AALTO, Finland*

*[2]Faculty of Pure and Applied Science, University of Tsukuba, Tsukuba 305-8573 Japan*

*[3]Department of Electrical Engineering and Computer Science, Massachusetts Institute of Technology, Cambridge, MA 02139 USA*

*E-mail: jori.lemettinen@aalto.fi

We present the electrical characterization of N-polar AlN layers grown by metal-organic vapor phase epitaxy and the demonstration of N-polar AlN-channel metal-semiconductor field-effect transistors (MESFETs). A high concentration of silicon is unintentionally incorporated during the high-temperature growth of N-polar AlN, causing high buffer leak current. The silicon concentration reduces from $2\times10^{18}/cm^3$ to $9\times10^{15}/cm^3$ with decreasing growth temperature, reducing the buffer leak current to 5.6 nA/mm at a 100 V bias. The N-polar AlN MESFET exhibits an off-state drain current of 0.27 nA/mm and a transistor on/off ratio of $4.6\times10^4$ due to the high-resistivity AlN buffer layers.

Nitrogen-polar (N-polar) III-nitride (III-N) semiconductors have received much attention for light emitting and electronics applications.[1,2] In transistor applications, the N-polar high electron mobility transistors (HEMTs) have a reversed structure of GaN/AlGaN layers compared to incumbent metal-polar AlGaN/GaN HEMTs.[3,4]

The N-polar GaN/AlGaN HEMTs provide a low contact resistance with the top GaN layer and a high carrier confinement by natural back barrier in addition to a low leak current due to the AlGaN buffer layer.[5–7] The N-polar III-N HEMTs can be improved by increasing the Al-content in the AlGaN buffer layer. The high Al-content AlGaN buffer layers have a low buffer leak current and induce higher carrier concentrations. In the extreme case, N-polar AlN buffer layers are the ideal transistor platform, which offers the highest back barrier and the highest thermal conductivity in III-Ns and low buffer leak current.[8–10]

However, it has been difficult to grow N-polar III-Ns with high crystalline quality. N-polar III-Ns have exhibited inversion domains, hexagonal hillocks and step bunching.[11–13] Recently, we grew a smooth N-polar AlN layer on SiC by optimizing the growth conditions in the high temperature regime using metal-organic vapor phase epitaxy (MOVPE).[14] A high-resistivity buffer layer with high crystalline quality and low impurity concentrations is a key for high-performance AlN-based transistors. In this letter, we report on the impurity concentration in N-polar AlN layers grown at various temperatures in addition to the demonstration of N-polar AlN channel metal-semiconductor field-effect transistors (MESFET).

N-polar AlN layers were grown on semi-insulating C-face SiC on-axis substrates by MOVPE.[13–15] The SiC surfaces were cleaned *in-situ* for 20 min at a nominal substrate surface temperature of 1180 °C in hydrogen ($H_2$) ambient at 100 mbar pressure prior to AlN growth. The substrate temperature was monitored using an emissivity corrected *in-situ* pyrometer. All temperature values in this work refer to the pyrometer estimated substrate temperature. The MOVPE reactor had a close coupled showerhead configuration. Trimethylaluminium (TMAl), ammonia ($NH_3$) and disilane ($Si_2H_6$) were used as precursors for aluminum, nitrogen and silicon, respectively. AlN was grown at 50 mbar pressure with $H_2$ as the carrier gas. In this work, we prepared multi-layer AlN stacks with different growth conditions to investigate the relation between the MOVPE growth conditions and the impurity incorporation into N-polar AlN layers. We grew first a 400-nm-thick unintentionally doped (UID) AlN layer, which acted as a marker layer, at 1150 °C with a V/III ratio of 20000 to obtain a smooth surface. After the first layer, we grew pairs of marker layers and 200-nm-thick UID AlN sample layers with different growth conditions.[14,15] The growth rates for the marker layers and the sample layers were 0.1 µm/h and between 0.3 µm/h to 0.4 µm/h, respectively. Secondary-ion mass spectroscopy (SIMS) measurements, which were performed at EAG laboratories, were used to determine the impurity concentrations in the AlN layers as a function of depth.

Figure 1 presents the impurity concentrations of the UID N-polar AlN layers grown with various V/III ratios between 1000 and 10000. The growth temperature was maintained at 1150 °C. The oxygen, hydrogen and carbon concentrations were $3\times10^{17}$/cm$^3$, $1\times10^{17}$/cm$^3$, $3\times10^{16}$/cm$^3$, which were similar or slightly lower than the reported values for high-temperature Al-polar AlN and N-polar GaN.[16,17] The concentration of these impurities was independent of the V/III ratio in this growth regime. However, the silicon concentration significantly depended on the growth conditions. The silicon concentration was increased by decreasing the ratio V/III to 1000, reaching a level comparable to the marker layers ($2\times10^{18}$/cm$^3$). We consider that the high concentration of silicon in AlN layers grown with a low V/III ratio of 1000 is caused by degraded crystalline quality.[14] The silicon concentration in N-polar AlN layers grown using V/III ratios of 10000 and 5000 is reduced by a factor of five compared to the marker layers. There is little difference in silicon concentration between the sample layers grown with the V/III ratio of 10000 and 5000. We suggest that the reduction of silicon concentration may result from the increased growth rate because the growth rate was roughly quadrupled compared to the marker layer conditions. The growth regime of high V/III ratio together with the increased growth rate is preferable for low unintentional silicon incorporation.

Figure 2 presents the impurity concentrations in UID N-polar AlN layers grown at various temperatures between 1050 °C and 1110 °C. The V/III ratio was 8000 for all the sample layers. The incorporation of oxygen, hydrogen and carbon impurities were independent of the growth temperature in this growth regime. However, the silicon concentration significantly depended on the growth temperature. The silicon concentration in the marker layers was $2\times10^{18}$/cm$^3$, which is in the same range as for high-temperature Al-polar AlN.[16] On the other hand, the sample layer grown at 1050 °C had a much lower silicon concentration of $9\times10^{15}$/cm$^3$, which is close to N-polar GaN and low-temperature Al-polar AlN.[17,19] We believe that the unintentional silicon incorporation into the AlN layers is derived from the etching of the SiC-coated susceptor in the $H_2$/$NH_3$ atmosphere.[18,19] The independence of carbon concentration is most likely due to efficient mechanisms for carbon surface removal or suppression of incorporation.[18]

Three types of samples were prepared to investigate the leak current through the N-polar AlN buffer layers. As schematically shown in Fig. 3 inset, all samples had a 50-nm-thick UID AlN layer grown at V/III ratio of 20000 and 1150 °C to reduce threading dislocation density (TDD), a 10-nm-thick UID AlN layer to obtain a smooth surface, and an intentionally doped 100-nm-thick n$^{++}$-AlN layer with nominal silicon doping density of 6×10$^{18}$/cm$^3$ to reduce contact resistance (growth rate 0.1 μm/h). Between the 50-nm-thick and 10-nm-thick UID AlN layers, three types of 400-nm-thick AlN buffer layers were inserted with different growth conditions; V/III ratio of 20000 and temperature of 1150 °C (growth rate 0.1 μm/h), V/III ratio of 10000 and temperature of 1100 °C (growth rate 0.3 μm/h), and V/III ratio of 10000 and temperature of 1050 °C (growth rate 0.3 μm/h). A low TDD is important to reduce the buffer leak current.[20] To investigate the crystalline quality, a separate 200-nm-thick UID AlN layer was measured using X-ray diffraction. The measured rocking curve full width half maximums were 120 arcsec and 210 arcsec for (002) and (102) reflections, respectively. Based on our previous results, we estimate that order of magnitude for the TDD was $10^8 - 10^9$/cm$^2$.[15]

Two contacts separated by a 4 μm spacing were formed for the electrical measurements. A 200-nm-deep Cl$_2$-based reactive-ion etching was performed to electrically isolate the electrodes to analyze the resistivity of the 400-nm-thick UID AlN buffer layer. A Ti (20 nm)/Al (100 nm)/ Ni (25 nm)/ Au (50 nm) metal stack was deposited using electron-beam evaporation for electrical contacts, followed by annealing at 800 °C for 30 s in a nitrogen ambient to form a metal alloy. Agilent B1505A power-device analyzer and a high-voltage Tesla probe station were used for the measurements.

Figure 3 shows the leak current through the UID AlN buffer layers as a function of applied voltage between the electrodes. All samples exhibited a leak current less than 0.3 nA/mm at the applied voltage of 20 V. The leak current increased with increasing growth temperature of the AlN buffer layer. The leak currents through the 1100 °C-buffer and 1050 °C-buffer layer were still less than 1 nA/mm at 60 V and 100 V bias, respectively, while the leak current thorough the 1150 °C-buffer layer dramatically increased at the same bias. We consider that the high leak current through the UID AlN buffer layers grown at high temperature resulted from the high concentration of unintentionally incorporated silicon with an estimated concentration of 2×10$^{18}$/cm$^3$. Further high resistivity of the N-polar AlN-buffer layers would be achieved by employing extrinsic compensation doping, such as iron or carbon, and/or by changing the susceptor coating material to prevent unintentional silicon incorporation.[19,21,22]

    N-polar AlN channel MESFETs with the UID AlN buffer layer grown at 1050 °C were fabricated, as schematically shown in Fig. 4. A Ni (30 nm)/ Au (200 nm) metal stack was deposited for the gate electrode in addition to the above described processing of mesa isolation and source/drain contact metallization. The MESFETs had a gate length ($L_g$) of 2 μm and gate-source spacing ($L_{gs}$) of 8 μm. The electron mobility and concentration estimated using Hall-effect measurement were 83 cm$^2$/(Vs) and 1×10$^{15}$/cm$^3$ at room temperature, respectively, which are close to a reported Al-polar n$^{++}$-AlN layer.[23] The electron concentration is low compared to the nominal doping density (6×10$^{18}$/cm$^3$) due to the low electrical-activation ratio of AlN.[23]

    Figure 5(a) presents the output characteristics of the N-polar AlN MESFET at room temperature. The AlN MESFETs have a normally-on operation with a pinch-off gate voltage of ($V_{gs}$) < -17 V. Drain current ($I_d$) is effectively modulated by $V_{gs}$ and shows good saturation. The maximum $I_d$ was 12.4 μA/mm for $V_{gs}$ = +1V, which is similar to an Al-polar AlN MESFET[8]. The subthreshold swing was 2.1 V/decade.

    Figure 5(b) presents the transfer characteristics of the N-polar AlN MESFET at a drain-voltage ($V_{ds}$) = +10 V at room temperature. The off-state $I_d$ was 0.27 nA/mm. The on/off ratio of the AlN MESFET was 4.6×10$^4$ which was mainly limited by the maximum $I_d$. The low maximum $I_d$ was attributed to both high contact and sheet resistances. Heavy doping in n$^{++}$-AlN contact layer and optimization of the metal alloy used in source-drain contacts would reduce the contact resistance. The improvement of the crystalline quality of N-polar AlN layers would reduce the sheet resistance. N-polar AlN channel MESFETs still have a lot of room to improve the device performance.

    Figure 6(a) presents three-terminal breakdown voltage ($V_{br}$) of the N-polar AlN MESFETs as a function of drain-gate spacing ($L_{dg}$) at room temperature. $V_{br}$ increased linearly with increasing drain-to-gate spacing $L_{dg}$. The effective critical electric field ($E_c$) of the three terminal $V_{br}$ is 0.36 MV/cm. Although the theoretical $E_c$ of AlN is 12 MV/cm, which offers promise for high-performance vertical AlN devices. Effective $E_c$ of 1 MV/cm and 0.81 MV/cm have been reported for lateral Al-polar AlN MESFET and Al-polar AlN/AlGaN HEMT which employed much thicker buffer layers (1 μm and 2 μm), respectively.[8,24] The further optimization of N-polar AlN growth conditions in addition to a field plate would significantly improve the breakdown characteristics.[25] Fig. 6(b) presents the three-terminal off-state breakdown characteristics of the N-polar AlN MESFET with $L_{dg}$ of 30 μm and gate voltage of -10 V. The maximum three-terminal $V_{br}$ was 1180 V when

the $L_{dg}$ was 30 µm. Breakdown voltages of 810 V, 1650 V and 2370 V have been reported for an Al-polar AlN/AlGaN HEMT, an $Al_{53}Ga_{47}N/Al_{38}Ga_{62}N$ HEMT and an AlN MESFET, respectively.[8,24,25] The N-polar AlN MESFET achieves good breakdown characteristics with a much thinner epitaxial stack. The off-state $I_d$ was less than 10 nA/mm at -1000 V. This is three orders magnitude lower than the reported off-state $I_d$ of 40 µA/mm for an $Al_{53}Ga_{47}N/Al_{38}Ga_{62}N$ HEMT at comparable voltages and similar to an Al-polar AlN MESFET.[8,25] These result show the potential of the N-polar AlN buffer for high-power applications.

In summary, we investigated the impurity concentration in N-polar AlN layers with various growth conditions and demonstrated the N-polar AlN-channel MESFETs. The oxygen, hydrogen, carbon and silicon concentrations of optimized low-temperature 1050 °C grown N-polar AlN were $3\times10^{17}/cm^3$, $1\times10^{17}/cm^3$, $3\times10^{16}/cm^3$ and $9\times10^{15}/cm^3$, respectively. The silicon concentration increased with increasing growth temperature. High-temperature UID AlN layer grown at 1150 °C exhibited a high silicon concentration of $2\times10^{18}/cm^3$. The good insulation of N-polar AlN buffer layers was achieved mainly by ensuring low silicon concentration at a reduced growth temperature. Using the optimized AlN buffer, an N-polar AlN channel MESFET with low leak current was demonstrated. The off-state drain current was 10 nA/mm at -1000 V which is three orders magnitude lower than that of an $Al_{38}Ga_{62}N$ buffer.

The authors acknowledge the funding support from the Academy of Finland (grant 297916) and JSPS KAKENHI Grant No. 16H06424 and 17K14110. A part of the research was performed at the OtaNano - Micronova Nanofabrication Centre of Aalto University. This work was carried out in part through the use of MITs Microsystems Technology Laboratories.

**Acknowledgments**

The authors acknowledge the funding support from the Academy of Finland (Grant 297916) and JSPS KAKENHI Grant No. 16H06424 and 17K14110. A part of the research was performed at the OtaNano - Micronova Nanofabrication Centre of Aalto University. This work was carried out in part through the use of MITs Microsystems Technology Laboratories.

**Figure Captions**

**Fig. 1.** SIMS depth profiles of silicon (solid black), oxygen (solid red), hydrogen (dashed black) and carbon (dashed red) concentrations in N-polar AlN layers grown with various V/III ratios at growth temperature of 1150 °C. The V/III ratios of the sample layers were 1000, 5000, and 10000. The marker layers and the sample layers are indicated by the white and gray areas, respectively. The marker layers were grown with V/III ratio of 20000 at growth temperature of 1150 °C.

**Fig. 2.** SIMS depth profiles of silicon (solid black), oxygen (solid red), hydrogen (dashed black) and carbon (dashed red) concentrations in N-polar AlN layers grown with various growth temperatures at V/III ratio of 8000. The growth temperatures of sample layers were 1050, 1085, and 1110 °C. The marker layers and the sample layers are indicated by white and gray areas, respectively. The marker layers were grown at V/III ratio of 20000 at a growth temperature of 1150 °C.

**Fig. 3.** Current-voltage characteristics through three types of N-polar AlN buffer layers with a two-terminal structure. The growth temperatures of the used AlN buffer layers are 1150 °C, 1100 °C and 1050 °C.

**Fig. 4.** Schematic cross-sectional view of the N-polar n-type AlN-channel MESFET with a buffer layer with UID-AlN/ high-resistivity AlN grown at 1050 °C/ high-quality AlN stacks.

**Fig. 5.** (a) N-polar AlN MESFET transfer characteristics at a drain voltage of +10 V with a gate length of 2 μm for $V_{gs}$. (b) DC output characteristics for $V_{gs}$ from +1 to -10 V.

**Fig. 6.** (a) Off-state breakdown voltage of the N-polar AlN MESFET as a function of gate-to-drain spacing. (b) Three-terminal off-state breakdown characteristics of the N-polar AlN MESFET with drain-to-gate spacing of 30 μm and gate voltage of -10 V.

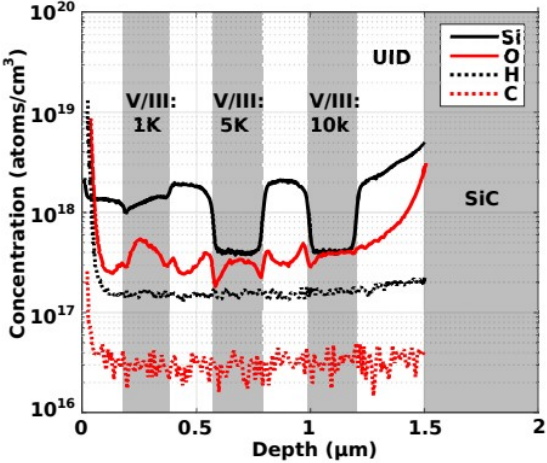

**Fig. 1.**

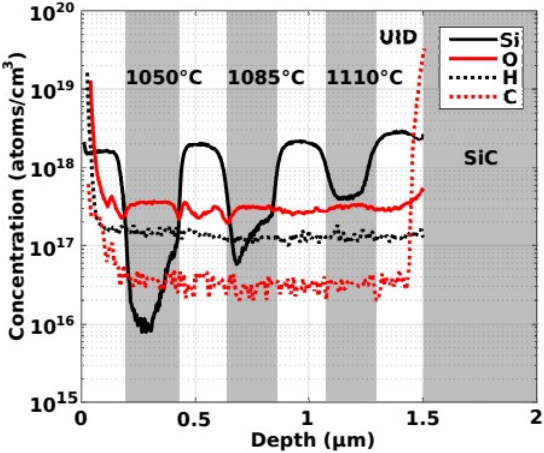

**Fig. 2.**

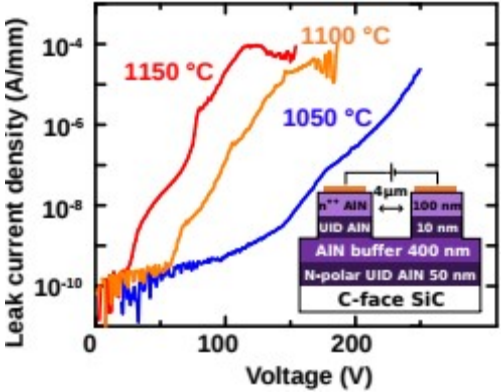

**Fig. 3.**

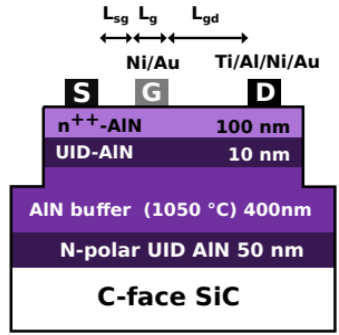

**Fig. 4.**

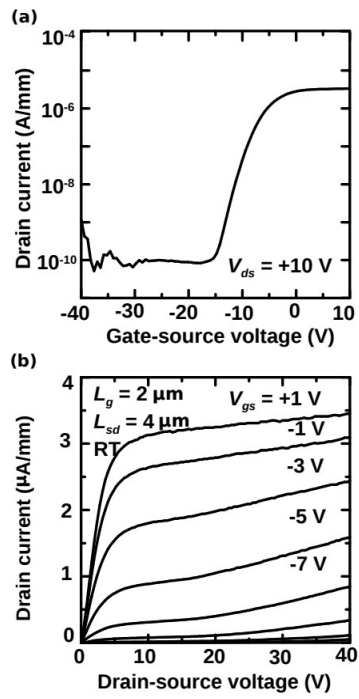

**Fig. 5.**

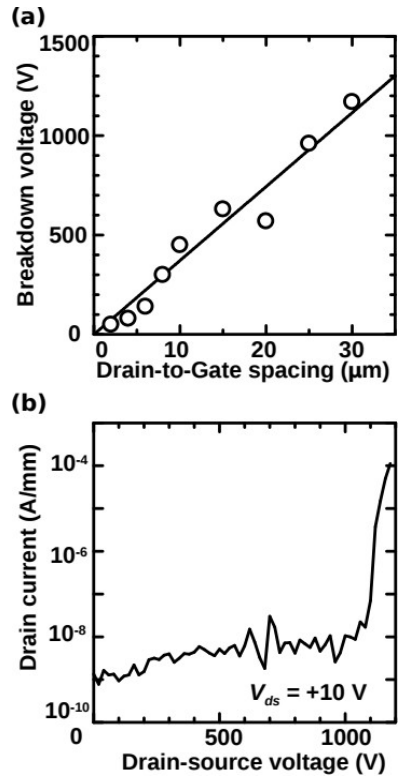

**Fig. 6.**